# A Reinforcement Learning Method For Power Suppliers' Strategic Bidding with Insufficient Information

Qiangang Jia, Zhaoyu Hu, Yiyan Li,
Zheng Yan and Sijie Chen
Key Laboratory of Control of Power
Transmission and Conversion, Ministry
of Education
Shanghai Jiao Tong University
Shanghai, 200240 China
jiaqiangang@sjtu.edu.cn

*Abstract*—Power suppliers can exercise market power to gain higher profit. However, this becomes difficult when external information is extremely rare. To get a promising performance in an extremely incomplete information market environment, a novel model-free reinforcement learning algorithm based on the Learning Automata (LA) is proposed in this paper. Besides, this paper analyses the rationality and convergence of the algorithm in case studies based on the Cournot market model.

*Keywords—reinforcement learning, learning automata, extremely incomplete information, Cournot market model*

## I. INTRODUCTION

### A. Backgrounds and Motivations

The power market is a typical imperfectly competitive environment for power suppliers. Several power suppliers can exercise market power to manipulate the market-clearing price and gain a higher profit. Besides, the market environment is like a black box for power suppliers. Thus many studies have applied reinforcement learning algorithms [1] to help power suppliers get optimal bidding strategies. Yu [2] and Kebriaei [3] used the Q learning to optimize the power suppliers' bidding strategies. Ye [4] applied the DDPG algorithm to help power suppliers bid.

Note that the optimizations like the above paper still need a lot of external information, e.g., the historical bids of the opponents, the parameters of the whole system. However, privacy protection has been paid more and more attention in today's society [5], which makes it more difficult for power companies to obtain external information. And sometimes the power suppliers can only get personal data, e.g., personal historical bids and profits. To this end, an algorithm suitable for this extremely incomplete information situation should be studied.

Learning Automata (LA) is a powerful tool to solve this problem. Learning automata is generally introduced in [6]. In [7], various learning automates are present, e.g., Finite Action Learning Automata (FALA) and Continuous Action Learning Automata (CALA). CALA is used widely in practical problems because the action space is continuous. In [8], CALA is used in the control of broadcast networks. And graph spectral partitioning and CALA are studied for supervised learning of large perceptual organization in [9]. The basic idea of CALA is to modify the probability distribution function (PDF) over the action space through iterations with the environment. These successful applications provide a bright prospect for the applications of this idea in the power market.

### B. Contributions and Scope

The contributions of this paper are outlined below:
- This paper proposes a novel reinforcement learning algorithm to help power suppliers optimize bids in an extremely incomplete information environment.
- Rationality and convergence [10] are tested in a Cournot duopoly model [11] without and with transmission line constraints.

### C. Paper Organizations

The rest of the paper is organized as follows. Section II presents the power suppliers' bidding procedure. Section III details the proposed algorithm. Section IV provides case studies to demonstrate the effectiveness of the algorithm. Section V discusses the conclusions and future extensions of this work.

## II. PROBLEM FORMULATION

### A. Market Structure.

The power market is composed of the power suppliers, the power consumers and the market operator.

The cost function of a single supplier can be written as

$$C = \frac{1}{2}mG^2 + nG + o \quad (1)$$

where $m$ and $n$ are the coefficient of the primary term and the secondary term, $o$ is the constant term. $G$ is its power output. In the Cournot market model, the supplier has motivations to change the bidding quantities to gain higher profit. Because there is a trade-off between the power output and the potential market clearing price: more power output will lead to a lower market-clearing price; less power output will lead to a higher market-clearing price.

The utility function of a single consumer can be written as

$$U = wD - \frac{1}{2}vD^2 \quad (2)$$

where $w$ and $v$ is the coefficient of the primary term and the secondary term, $D$ is the load demand of the consumer.

The market operator is to maximize social welfare by running the optimal power flow dispatch algorithm [12]. The constraints include the power balance constraint, the transmission line capacity constraints, etc.

## B. Bidding Procedure

In a single round of market clearing, suppliers bid their quantities strategically while consumers submit their utility functions to the market operators honestly. After the market is cleared, the market-clearing prices can be calculated and given to the corresponding supplier. And a single supplier can calculate the profit

$$P = \lambda G - C \quad (3)$$

where $P$ is the profit of the supplier, $\lambda$ is the locational marginal price (LMP) where the supplier is located. The objective of the supplier is to maximize the profit as (3).

## III. PROPOSED ALGORITHM FOR SUPPLIER'S STRATEGIC BIDDING

### A. Proposed Algorithm

The proposed algorithm is inspired by the CALA algorithm in [13]. The idea of the proposed algorithm is to modify the mean and the variance of the PDF over action space through interactions with the environment.

The action space in this bidding optimization problem under the Cournot market is the limits of bidding quantities. We define the PDF at iteration $t$ as a Gaussian Distribution $N(\mu_t, \sigma_t)$ (it can be replaced by other distributions according to actual needs). Each power supplier bids the random action $x$ and the mean action $\mu$ to the environment alternately. Set $b_x$ as the profit corresponding to the random action $x$, $b_\mu$ as the profit corresponding to the mean action $\mu$. The update rule of the proposed algorithm is demonstrated as follows

$$\mu_{t+1} = \mu_t + k_\mu \cdot \delta_\mu \quad (4)$$

$$\sigma_{t+1} = \sigma_t + k_\sigma \cdot \delta_\sigma - c \quad (5)$$

where $k_\mu$ is the update coefficient of the mean, $\delta_\mu$ is the update step of the mean, $k_\sigma$ is the update coefficient of the variance, $\delta_\sigma$ is the update step of the variance, $c$ is a small enough coefficient that gradually reduces the variance to stabilize the learning process, $t$ is the iteration time. The detailed definition of $k_\mu$ and $k_\sigma$ is as follows.

$$k_\mu = \begin{cases} sign(x-\mu), & \text{if } b_x > b_\mu \\ sign(\mu-x), & \text{if } b_\mu > b_x \end{cases} \quad (6)$$

$$k_\sigma = \begin{cases} 1 & \begin{array}{l} \text{if } b_x > b_\mu \ \& \ x-\mu > \sigma \text{ or} \\ b_x < b_\mu \ \& \ x-\mu <= \sigma \end{array} \\ -1 & \begin{array}{l} \text{if } b_x > b_\mu \ \& \ x-\mu < \sigma \text{ or} \\ b_x < b_\mu \ \& \ x-\mu >= \sigma \end{array} \end{cases} \quad (7)$$

If the action $x$ results in better profit than the action $\mu$, we move $\mu$ towards $x$; otherwise, we move $\mu$ away from $x$. And we increase $\sigma$ when an action leaves the mean by a standard deviation and results in a better reinforcement signal or when an action closes to the mean by a standard deviation and results in a worse reinforcement signal; otherwise, the $\sigma$ is decreased. Besides, the update step of mean and standard deviation is fixed to a constant value, making the learning process more robust and to achieve a better performance in convergence.

This algorithm only needs the supplier's personal data including the historical bids and profits. Then the learning process can be executed. This is workable in the extremely incomplete information environment.

The detailed process of the learning process is given in the pseudo-code.

| Proposed Algorithm |
|---|
| 1: Initialize the PDF $N(\mu_0, \sigma_0)$, define the iteration limit $M$; |
| 2: **for** $t$ = 1:$M$ **do** |
| 3:   **if** mod($t$,2) == 1 **then** |
| 4:     $a_t \leftarrow \mu$ |
| 5:   **else if** mod($t$,2) == 0 **then** |
| 6:     $a_t \leftarrow x$ randomly generated from $N(\mu_t, \sigma_t)$; |
| 7:   **end if** |
| 8:   Choose and submit $a_t$ to the market operator; |
| 9:   Gain profit $P_t$ by (3) after the market is cleared; |
| 10:   Save $t$, $a_t$ and $P_t$ as experience to the data buffer; |
| 11:   **if** $t > 2$ **then** |
| 12:     **if** mod($t$,2) == 1 **then** |
| 13:       $b_\mu \leftarrow P_t$ |
| 14:       $b_x \leftarrow P_{t-1}$ |
| 15:     **else if** mod($t$,2) == 0 **then** |
| 16:       $b_\mu \leftarrow P_{t-1}$ |
| 17:       $b_x \leftarrow P_t$ |
| 18:     **end if** |
| 19:     Perform the PDF update as (4)-(7); |
| 20:   **end if** |
| 21: **end for** |
| **End the Algorithm** |

### B. Rationality and Convergence

Criteria for evaluating the effectiveness of a multi-agent reinforcement learning algorithm are usually rationality and convergence.

Rationality means that if the other suppliers take the stationary policies then the learning algorithm will converge to a policy that is a best-response to their policies. Convergence means that the game will necessarily converge to a stationary policy when all the players are using the learning algorithm. These two indexes will be analyzed in the case study.

## IV. CASE STUDY

The simulations are run at Matlab2018b on a PC with a 16GB RAM. The main objective lies in validating its rationality and convergence. The topology of the testing system [14] is shown in Fig.1.

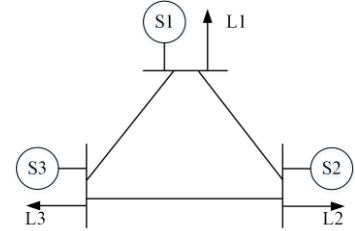

Fig. 1. 3-bus system topology.

There are three suppliers and three consumers in the system (the lower and upper power output limits are 0 and 2000MW for the three suppliers, and the power demands for consumers are unlimited). The parameters of them are shown in TABLE.I.

TABLE I. PARAMETERS OF THE 3-BUS SYSTEM

| Node | Utility Function | | Cost Function | | |
|---|---|---|---|---|---|
| | $w_i$ ($/MWh) | $v_i$ ($/MW²h) | $m_i$ ($/MW²h) | $n_i$ ($/MWh) | $o_i$ ($/h) |
| 1 | 108.4096 | 0.0555 | 0.015718 | 1.360575 | 9490 |
| 2 | 103.8283 | 0.066909 | 0.021052 | -2.07807 | 11128 |

| 3 | 105.6709 | 0.063703 | 0.012956 | 8.105354 | 6821 |

In the scenario without transmission line constraints, the line is unbounded; In the scenario with transmission line constraints, the capacity of line 1-3 is bounded to 16MW to cause congestion.

The parameters of the algorithm are as follows.

The iteration limit $M$ is 6000. $\delta_\mu$ and $\delta_\sigma$ are set to be 1 and 0.2 respectively. The initial mean and standard deviation are 600 and 20 respectively. The coefficient $c$ is $10^{-3}$.

And we will compare the learning results calculated by the proposed algorithm with the Nash equilibrium calculated by the analytical EPEC method [15] (under a global perspective for comparison, which is impossible in practice).

*a. Rationality of the algorithm.*

The rationality is judged from three criteria, e.g., the profit, the action, and the LMP. To test the rationality, the strategies of supplier 2 and supplier 3 are fixed to 1046MW and 995MW in the environment without transmission line constraints while 1268MW and 645MW in the environment with transmission line constraints. The learning results of supplier 1 are shown in TABLE. II and TABLE. III respectively (the action learned by the proposed algorithm is the mean of the PDF).

TABLE II. THE LEARNING RESULT WHEN ONLY SUPPLIER 1 LEARNS WITHOUT CONSTRAINTS

| Method | Supplier | Criterion | | |
|---|---|---|---|---|
| | | *Profit*($/h) | *Action*(MW) | *LMP*($/MWh) |
| Proposed Algorithm | 1 | 25231 | 1103 | 41.51 |
| Analytical EPEC | 1 | 25221 | 1105 | 41.45 |

TABLE III. THE LEARNING RESULT WHEN ONLY SUPPLIER 1 LEARNS WITH CONSTRAINTS

| Method | Supplier | Criterion | | |
|---|---|---|---|---|
| | | *Profit*($/h) | *Action*(MW) | *LMP*($/MWh) |
| Proposed Algorithm | 1 | 24779 | 791 | 51.43 |
| Analytical EPEC | 1 | 24306 | 781 | 50.77 |

From these two tables, we can find that the profits, the actions and the LMPs are 25231$/h, 1103MW and 41.51$/MWh without transmission line constraints and 24779$/h, 791MW and 51.43$/MWh with transmission line constraints. The percentage errors of the profits, the actions and LMPs are 0.04%, 0.18% and 0.14% without transmission line constraints and 1.9%, 1.3% and 1.3% with transmission line constraints compared to the best response calculated by the analytical EPEC method. The rationality of this algorithm is proved.

*b. Convergence of the algorithm.*

To test the convergence of the proposed algorithm, the two suppliers are all strategic players using this algorithm. The result is shown in TABLE.IV and TABLE.V in the scenarios without and with transmission line constraints respectively.

TABLE IV. THE RESULTS WHEN ALL SUPPLIERS LEARN WITHOUT CONSTRAINTS

| Method | Supplier | Criterion | | |
|---|---|---|---|---|
| | | *Profit*($/h) | *Action*(MW) | *LMP*($/MWh) |
| Proposed Algorithm | 1 | 25436 | 1109 | 41.57 |
| | 2 | 22771 | 1035 | |
| | 3 | 20103 | 997 | |
| Analytical EPEC | 1 | 25221 | 1105 | 41.45 |
| | 2 | 22891 | 1046 | |
| | 3 | 19947 | 995 | |

TABLE V. THE RESULTS WHEN ALL SUPPLIERS LEARN WITH CONSTRAINTS

| Method | Supplier | Criterion | | |
|---|---|---|---|---|
| | | *Profit*($/h) | *Action*(MW) | *LMP*($/MWh) |
| Proposed Algorithm | 1 | 25235 | 815 | 50.37 |
| | 2 | 38209 | 1261 | 50.32 |
| | 3 | 17476 | 639 | 50.27 |
| Analytical | 1 | 24306 | 781 | 50.77 |
| | 2 | 38949 | 1268 | 50.76 |
| | 3 | 17977 | 645 | 50.75 |

The percentage errors of the profits, the actions and the LMPs are shown in TABLE.VI.

TABLE VI. PERCENTAGE ERRORS IN DIFFERENT SCENARIOS

| Scenario | Supplier | Criterion | | |
|---|---|---|---|---|
| | | *Profit*(%) | *Action*(%) | *LMP*(%) |
| Without Constraints | 1 | 0.85 | 0.36 | 0.29 |
| | 2 | 0.52 | 1.1 | |
| | 3 | 0.78 | 0.20 | |
| With Constraints | 1 | 3.8 | 4.4 | 0.78 |
| | 2 | 1.9 | 0.55 | 0.86 |
| | 3 | 2.8 | 0.93 | 0.95 |

From these three tables, we can find that: without transmission line constraints, the profits of the three suppliers are 25436$/h, 22771$/h and 20103$/h, the actions are 1109MW, 1035MW and 997MW, the LMPs are 41.57$/MWh. With transmission line constraints, the profits of the three suppliers are 25235$/h, 38209$/h and 17476$/h, the actions are 815MW, 1261MW and 639MW, the LMPs are 50.37$/MWh, 50.32$/MWh and 50.27$/MWh. The percentage errors of the profits, the actions and LMPs are all within 4.4% compared to the Nash equilibrium calculated by the analytical EPEC method and this proves the convergence. The changing process of the profits, the actions as well as the LMPs are shown via Fig.2, Fig.3 and Fig4.

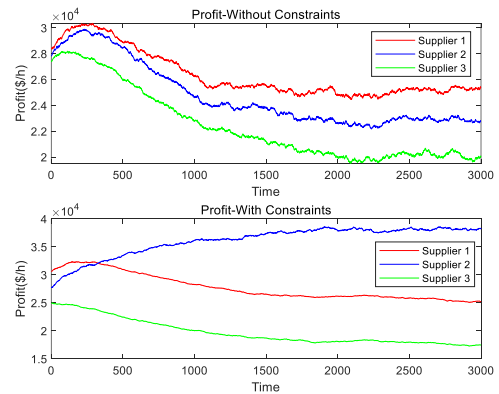

Fig. 2. Profit of suppliers in the 3-bus system.

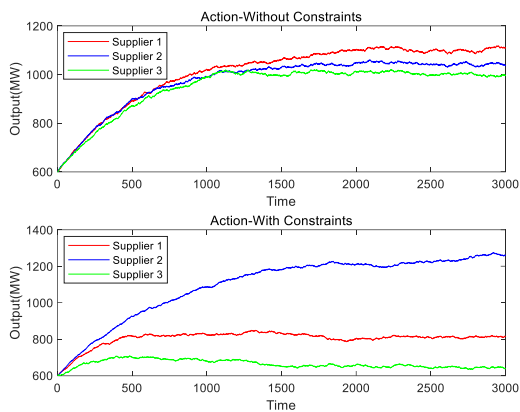

Fig. 3. Action of suppliers in the 3-bus system.

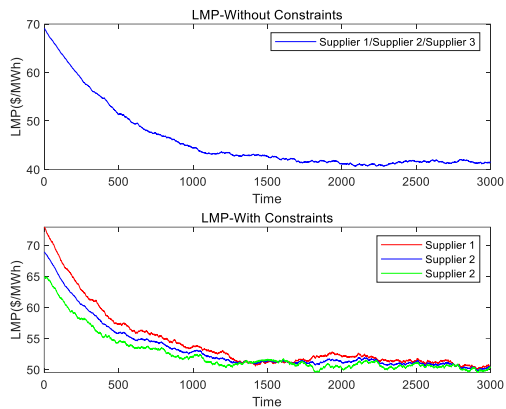

Fig. 4. LMP of suppliers in the 3-bus system.

## V. CONCLUSION AND FUTURE WORK

This paper proposes a novel learning algorithm to help power suppliers to optimize bids under extremely incomplete information. The simulation results show its rationality and convergence.

This algorithm is of great engineering and practical value to instruct the power suppliers to bid wisely. Next, we will consider extending the method to a real power network with more nodes and further improving the learning speed of the method.